\documentstyle[12pt]{article}
\begin{document}
\title{A family of multi-value cellular automaton model\\
for traffic flow}  
\author{
Katsuhiro Nishinari$^a$ and Daisuke Takahashi$^b$\\
$^a$ Department of Applied Mathematics and Informatics, \\
Ryukoku University, Seta, Ohtsu 520--2194, JAPAN\\
knishi@rins.ryukoku.ac.jp\\
$^b$ Department of Mathematical Sciences, Waseda University,\\
Ohkubo 3--4--1, Shinjuku-ku, Tokyo 169--8555, JAPAN \\
daisuke@mse.waseda.ac.jp\\} 
\date{}
\maketitle
\begin{abstract} 
In this paper, a family of multi-value cellular automaton (CA)
 associated with traffic flow is presented. 
The family is obtained by extending the rule--184 CA, which is an
 ultradiscrete analogue to the Burgers equation.  
CA models in the family show both metastable states and stop-and-go
 waves, which are often observed in real traffic flow.  
Metastable states in the models exist not only on a 
prominent part of a free phase but also in a congested phase.
\end{abstract}
\newpage
\section{Introduction}
Traffic phenomena attract much attention of physicists in recent years. 
It shows a complex phase transition from free to congested state, 
and many theoretical models have been proposed so
far\cite{Gr}--\cite{PN}.  
Among them we will focus on 
deterministic cellular automaton (CA) models.  
CA models are simple, flexible, and suitable for computer 
simulations of discrete phenomena.
\par
  The rule--184 CA\cite{Wo} has been widely used as a prototype of 
deterministic model of traffic flow. 
In the model, lane is single and cars can move by one site 
at most every time step.  
Its several variations have been proposed recently: 
First, Fukui and Ishibashi (FI) proposed a high 
speed extension\cite{FI} of the rule--184 CA. 
In the model, cars can move more than one site per unit time 
if there are successive vacant spaces ahead.  
Second, Fuk\'{s} and Boccara proposed 
a `monitored traffic model'\cite{FB}, 
which is a kind of quick start (QS) model. 
In the model, drivers can prospect vacant 
spaces due to car motion in the next site 
and can move more quickly compared with the rule--184 CA.  
Third, Takayasu and Takayasu proposed a slow start 
(SlS) model\cite{TT}, in which cars can not move 
just after they stop and wait for a unit time.  
This represents an asymmetry of stopping and starting behavior.  
We call all these variants a family of rule--184 CA in this paper.
\par
  A desirable condition for CA as a traffic model is that 
it can show a phase transition observed in real data.  
Let us see an example of real data of Tomei expressway 
taken by the Japan Highway Public Corporation\cite{NH}.  
Fig.\ref{fig1} shows flow--density diagrams often 
called `fundamental diagram', of both lanes in January 
1996 on up line at 170.64 km point from Tokyo.  
Fig.\ref{fig1} (a) and (b) are diagrams of driving and  
acceleration lane respectively. 
We see that a phase transition from free to congested state around a
density of 25 vehicles$/$km in both lanes, 
and there is a clearer discontinuity in the acceleration lane.  
The multiple states of flow at the same density around 
the critical density are also observed particularly 
in the acceleration lane, and a skeleton curve of 
plot points shows a shape of `inverse $\lambda$'.  
These phenomena are often observed in real data\cite{NH}.
\par
  There is another observed fact that over a certain critical density, 
a perturbation to a uniform traffic causes a formation of jam, 
and `stop-and-go' wave propagates backward\cite{HMPR}.  
Cars accelerate and decelerate alternately in the wave.  
We adopt these two experimental facts, multiple states 
and stop-and-go wave, as criteria to judge whether 
a CA is suitable for traffic model or not.  
Among the rule--184 CA family, only SlS model 
shows multiple states and also shows a stop-and-go wave.
\par
  Recently, a multi-value generalization of the rule--184 CA 
has been proposed by using an ultradiscrete method\cite{TTMS}. 
Its evolution equation is
\begin{equation}
  U_j^{t+1} = U_j^t + \min(U_{j-1}^t, L - U_j^t)
                    - \min(U_j^t, L - U_{j+1}^t),
\label{BCA}
\end{equation}
where $U^t_j$ represents the number of cars at site $j$ and time $t$, 
and $L$ is an integer constant.  Each site is assumed to 
hold $L$ cars at most.  
We can prove that if $L>0$ and $0\le U_j^t\le L$ for any $j$, 
then $0\le U_j^{t+1}\le L$ holds for any $j$.  
Thus (\ref{BCA}) is considered to be $(L+1)$--value CA.  
Since (\ref{BCA}) is obtained from an ultradiscretization of 
the Burgers equation, we call it Burgers CA (BCA)\cite{NT1}.  
Note that BCA is equivalent to the rule--184 CA in a special case of
$L=1$. 
The positive integer $L$ can be interpreted physically 
in the following three ways: 
First, the road is an $L$--lane freeway in a coarse sense, 
and effect of lane-changing rule is not expressed explicitly.  
In Appendix, we show BCA with $L=2$ can be interpreted as 
a two lane model of the rule--184 CA with an explicit lane--changing
rule.  
Second, $U_j^t/L$ represents a probability distribution of 
cars in a single-lane freeway.  
In this case, $U_j^t$ itself no longer represents 
a real number of cars at site $j$.  
Third, also in a single-lane freeway, length of each 
site is assumed to be long enough to contain $L$ cars.  
By introducing the free parameter $L$ into the rule--184 CA family, 
we can generalize them to multi-value CA and obtain rich algebraic 
properties and wide applications to various transport phenomena.
\par
  Though BCA does not show multiple states at a density 
in the fundamental diagram, we have shown that its high 
speed extension shows multiple states around a critical
density\cite{NT2}.  
Metastable states of flow in the model exist on a prominent 
part of a line in a free phase over a critical density.  
The model is also considered to be a multi-value 
generalization of FI model.  
This model is mentioned in Sec.\ref{sec:EBCA2} to compare other models.

Other models in the rule--184 CA family, SlS model and QS model, 
are two-value CA like FI model.  We generalize them to multi-value 
ones and investigate their properties as a traffic model in detail.
\par
  We use a max--plus representation to express time evolution 
rule of those generalized models.  This representation comes 
from the ultradiscrete method which has a close relation with 
the max--plus algebra\cite{maxplus}, as shown in our previous papers.  
Using max--plus representation, we can express evolution 
rule of the above CA's in a conservation from like (\ref{BCA}), 
which automatically derives a conservation of total number of cars.
\section{Generalization to multi-value CA}  \label{sec:GMCA}
In this section, we will present four CA models which are 
multi-value generalization of the rule--184 CA family.  
Before explaining models, we shortly review a rule of car movement of
BCA.  
Note that we assume $L$ is a lane number of the road in the 
description of all models.  In BCA, cars at site $j$ move to 
vacant spaces at their next site $j+1$ as many as possible.  
Therefore, car flow from $j$ to $j+1$ at time $t$ 
becomes $\min(U_j^t, L-U_{j+1}^t)$.  
Thus adding $\min(U_{j-1}^t, L-U_j^t)$ (flow from $j-1$ to $j$) 
and $-\min(U_j^t, L-U_{j+1}^t)$ (flow from $j$ to $j+1$) 
to $U_j^t$ (present car number), we obtain $U_j^{t+1}$ and derive (\ref{BCA}).
\subsection{Multi-value QS model}
First, we generalize QS model to a multi-value one.  
In the multi-value model, cars at site $j$ move to vacant spaces at 
site $j+1$ per unit time as many as possible.  
This movement rule is similar to that of BCA.  
However, a difference is that drivers at $j$ estimate 
vacant spaces at $j+1$ by predicting how many cars move from $j+1$ to
$j+2$.  
Thus, they estimate vacant spaces at $j+1$ at time $t$ to 
be $L-U_{j+1}^t \hbox{(present spaces)} + 
\min(U_{j+1}^t,L-U_{j+2}^t) \hbox{(predicted spaces due to BCA
movement)}$.  
Therefore considering the number of cars coming into and 
escaping from site $j$, evolution equation on $U_j^t$ is given by
\begin{eqnarray}
 U^{t+1}_j
&=& U^t_j
    + \min(U^t_{j-1}, L-U^{t}_{j} + \min (U^{t}_{j}, L-U^{t}_{j+1}))
                                                               \nonumber\\
& & - \min(U^t_{j}, L-U^{t}_{j+1} + \min (U^{t}_{j+1}, L-U^{t}_{j+2}))
                                                               \nonumber\\
&=& U^t_j
    + \min (U^t_{j-1}, 2L-U^{t}_{j}-U^{t}_{j+1})
    - \min (U^t_{j}, 2L-U^{t}_{j+1}-U^{t}_{j+2}).
\label{spsca}
\end{eqnarray}
The final form of (\ref{spsca}) means that the number of 
movable cars ($U_j^t$) is limited by vacant spaces 
($2L-U_{j+1}^t-U_{j+2}^t$) in their next two sites.  
In the case of $L=1$, this model is identical to QS model proposed 
by Fuk\'{s} and Boccara, and it is the rule--3212885888 CA of 
a neighborhood size `5' after Wolfram's terminology\cite{Wo}.
\subsection{Multi-value SlS model}
In the original SlS model, lane is single and cars can not move 
just after they stop and wait for a unit time.  
Movable cars move to vacant spaces in their next site like BCA.  
In the multi-value case, we should distinguish standing cars 
and moving cars in each site because each site can hold plural cars.  
The number of cars at site $j$ blocked by cars at $j+1$ 
at time $t-1$ is represented by
$U^{t-1}_j - \min (U^{t-1}_j, L-U^{t-1}_{j+1})$.  
These cars cannot move at $t$, 
then the maximum number of cars movable to site $j$ at $t$ 
is given by $U^t_j - \{U^{t-1}_j - \min (U^{t-1}_j, L-U^{t-1}_{j+1})\}$.
Therefore, the multi-value SlS CA is given by
\begin{eqnarray}
 U^{t+1}_j = U^t_j
&+& \min\left[U^t_{j-1} - \{U^{t-1}_{j-1} - \min(U^{t-1}_{j-1}, L-U^{t-1}_j)\},
              L-U^t_j \right]\nonumber\\
&-& \min\left[U^t_{j} - \{U^{t-1}_{j} - \min (U^{t-1}_{j}, L-U^{t-1}_{j+1})\},
              L-U^t_{j+1}) \right]
\label{SLS}
\end{eqnarray}
We note that this model includes the original SlS model in the case
$L=1$.  
Since $U^{t+1}$ is determined by $U^t$ and $U^{t-1}$, 
the evolution equation is second order in time and an 
effect of inertia of cars is included in this model.
\subsection{EBCA2 model} \label{sec:EBCA2}
In our previous paper\cite{NT2}, we propose an extended BCA (EBCA) model 
in which cars can move two site forward at a unit time 
if the successive two sites are not fully occupied.  
In this paper, we call the model `EBCA2' and 
another model with a similar extension `EBCA1' described 
in the next subsection.  
A main difference between two models is a priority of movement 
of fast and slow cars to vacant spaces.  
Fast cars with speed 2 move prior to slow ones with 
speed 1 in EBCA2 and otherwise in EBCA1.
\par
  Evolution equation of EBCA2 is given by
\begin{equation} 
U_j^{t+1} = U_j^t + \min(b_{j-1}^t+a_{j-2}^t, L-U_{j}^t+a_{j-1}^t)
                  - \min(b_{j}^t+a_{j-1}^t, L-U_{j+1}^t+a_{j}^t),
\label{EBCA2}
\end{equation}
where $a_j^t \equiv \min(U_j^t, L-U_{j+1}^t, L-U_{j+2}^t)$ 
which is the number of cars moving by two sites and 
$b_j^t \equiv \min(U_j^t,L-U_{j+1}^t)$.  
In reference \cite{NT2}, we only study the cases of $L=1$ and 2. 
In this paper, we show some properties of flow for $L$ in the next
section.  
Note that (\ref{EBCA2}) in the case of $L=1$ is FI model which is 
equivalent to the rule--3436170432 CA.
\subsection{EBCA1 model}
There is another possibility when we extend BCA to a high-speed 
model and we call this extended new model EBCA1.  
In contrast to EBCA2, slow cars with speed 1 move prior 
to fast ones with speed 2.  
Then car movement from $t$ to $t+1$ consists of the 
following two successive procedures,
\begin{itemize}
 \item[a)] Cars move to vacant spaces in their next site as many as possible.
 \item[b)] Only cars moved in procedure a) can move more one site and they
move to their next vacant spaces as many as possible.
\end{itemize}
The number of moving cars at site $j$ and time $t$ in procedure a) is 
given by $b_j^t \equiv \min(U_j^t,L-U_{j+1}^t)$.  
In procedure b), the number of moving cars at site $j+1$ 
becomes $\min(b_j^t, L-U^t_{j+2}-b^t_{j+1}+b^t_{j+2})$, 
where the second term in $\min$ represents vacant spaces at site $j+2$ 
after the first procedure a).  
Therefore, considering a total number of cars entering into and 
escaping from site $j$, evolution equation of EBCA1 is given by
\begin{eqnarray} 
  U_j^{t+1} &=& U_j^t + b^t_{j-1} - b^t_{j} \nonumber \\
            & &  + \min(b_{j-2}^t, L-U_{j}^t-b_{j-1}^t+b^t_{j})
                 - \min(b_{j-1}^t, L-U_{j+1}^t-b_{j}^t+b^t_{j+1}) \nonumber \\
            &=& U_j^t + \min(b_{j-1}^t+b_{j-2}^t, L-U_{j}^t+b_{j}^t)
                      - \min(b_{j}^t+b_{j-1}^t,L-U_{j+1}^t+b_{j+1}^t).
\label{EBCA1}
\end{eqnarray}
Note that (\ref{EBCA1}) with $L=1$ differs from FI model and 
it is equivalent to rule--3372206272 CA.
\par
  We can consider that this rule is an extension of SlS model to 
high-speed motion since cars which stop at procedure a) 
cannot move at procedure b).  
Let us consider a relation between SlS and EBCA1 models in detail.  
If we write each procedure of EBCA1 explicitly using an intermediate
time step, we obtain
\begin{eqnarray}
  U_j^{t+1/2} &=& U_j^t + b_{j-1}^t - b_j^t \label{EBCA1eq1}\\
  U_j^{t+1}   &=& U_j^{t+1/2}
                   + \min(U_{j-1}^{t+1/2}-\{U_{j-1}^t - b_{j-1}^t\},\,
                          L-U_j^{t+1/2}) \nonumber \\
              & & \qquad -\min(U_j^{t+1/2}-\{U_j^t - b_j^t\},\,
                          L-U_{j+1}^{t+1/2}). \label{EBCA1eq2}
\end{eqnarray}
Equations (\ref{EBCA1eq1}) and (\ref{EBCA1eq2}) represent above 
procedure a) and b) respectively, and $U_j^{t+1/2}$ denotes 
car number at site $j$ just after procedure a).  
If we replace $t+1$ by $t+1/2$ in (\ref{BCA}), we obtain
(\ref{EBCA1eq1}).  
Moreover, if we replace $t$ by $t+1/2$ and $t-1$ by $t$ in (\ref{SLS}), 
we obtain (\ref{EBCA1eq1}).  
Therefore we can consider EBCA1 to be a `combination' of BCA and SlS rules.
\section{Fundamental diagram and multiple states}
We discuss fundamental diagrams of new CA models described 
in the previous section.  
In the followings, we will consider a periodic road or a circuit.  
All models in Sec.\ref{sec:GMCA} can be expressed in a conservation form such as
\begin{equation}
 \Delta_t\,U^t_j + \Delta_j\,q^t_j = 0,
\end{equation} 
where $\Delta_t$ and $\Delta_j$ are forward difference operator 
with respect to indicated variable, and $q^t_j$ represents a traffic
flow.  
Average density $\rho$ and average flow $Q^t$ over all sites are defined by
\begin{equation}
  \rho \equiv {1\over KL} \sum_{j=1}^K U_j^t,\qquad
  Q^t  \equiv {1\over KL} \sum_{j=1}^K q^t_j,
\end{equation}
where $K$ is number of sites in a period.  
Since all models are in a conservation form, average density 
does not depend on time and we can use $\rho$ without a script $t$.
\par
  Figures \ref{fig2} (a)--(d) are density--flow diagrams of previous
  models 
with $L=2$ and $K=30$.  
Figures \ref{fig2} (a)--(d) corresponds to multi-value QS, 
multi-value SlS, EBCA2 and EBCA1 models respectively.  
If we set a number $N$ of cars, 
we obtain a unique density $\rho=N/KL$ but there are many 
initial distributions of cars.  
Each initial distribution does not always reach a steady flow and 
often make a periodic state.  $Q^t$ can also changes periodically in
time.  
Therefore, we plot every points in figures by averaging $Q^t$ from 
$t=2K$ to $t=4K$.  (We use $Q$ as an average flow.)  
Note that $t=2K$ is long enough from initial time to obtain 
a periodic state and $2K$ time steps is much longer than its period.
\par
  Moreover, we obtain different values of $Q$ from different 
initial distributions for a given $\rho$ in Figs.\ref{fig4} (b) $\sim$ (d).  
In Ref.\cite{NT2}, this type of state giving different $Q$ values 
for the same density are called `multiple states'\cite{NT2}.  
We use this terminology for other models in this paper.  
Multiple states exist around a critical density.  
Especially for multi-value SlS model (Fig.\ref{fig2} (b)) and EBCA1
model (Fig.\ref{fig2} (d)), 
we see a thick branch other than lines forming an inverse $\lambda$
shape and the distribution of data in the branch look random.  
This fact is interesting because evolution rules are completely
deterministic.  

Next, we focus on EBCA1 model and examine its properties in detail.  
Figure \ref{fig3} is a skeleton diagram of EBCA1 model.  
Branches in Fig.\ref{fig3} are drawn as straight lines and 
we obtain them using specific initial conditions.  For example, states 
`$\cdots 11102021110202 \cdots$', `$\cdots 121212 \cdots$',
`$\cdots 002002 \cdots$' and `$\cdots 222222 \cdots$'
are all steady states in the model, and 
flows of these states are plotted on B, C, D and E respectively 
in Fig.\ref{fig3}.  There are two branches in congested phase 
of higher density, that is, B--C and D--E.  
We call the branch D--E and B--C congested phase zero and phase one respectively.
It is noted that there are many states not on the skeleton in Fig.\ref{fig3}.  
For example, state $\cdots 211211 \cdots$ is a steady state of the
model, and it has density 2/3 and flow 1/2. 
 
Fig.\ref{fig4} shows a time evolution of flow $Q^t$ in EBCA1 model 
with a density of phase transition region.  
Number of total sites $K$ is 60 and 240 in Fig.\ref{fig4} (a) and (b) 
respectively.  
In the figures, we can see periodic oscillation 
of flow appearing soon after an initial time.  
We observe that the maximum period of oscillation becomes longer as 
system size becomes larger.  
Moreover, oscillation in a period is not simple as shown in the figures.
Fig.\ref{fig4} (c) shows a power spectrum $I$ versus frequency $f$
defined by $I=|\sum_{t=0}^{T-1}Q^t\exp(2\pi i ft/T)|/T$. 
We set $K=240$ and $T=1000$.
This indicates that the irregular oscillation is similar to white noise. 
\par
  Fig.\ref{fig5} is a fundamental diagram of EBCA1 with $L=1$.  
It is interesting that there exist multiple states even in the case
$L=1$, 
which is equivalent to the rule--3372206272 CA.  
Among deterministic two-value CA models, only SlS model is 
known to show metastable state so far.  
Thus, EBCA1 with $L=1$ becomes the second example of such kind of two-value CA.
\par
  At the end of this section, we give a comment on a case of $L>3$.  
Fundamental diagram of multi-value QS model is the same regardless of
$L$.  
In other models, small branches increases around 
the critical density as $L$ becomes larger.  
Let us show this phenomenon using EBCA2 which has clear branches.  
The fundamental diagram of EBCA2 with $L=7$ is given in Fig.\ref{fig6}.
There are many branches around the critical density. 
We can give a partial explanation of these branches.  
Let us assume initial values on all sites are restricted to $n$ or $L-n$ 
where $0\le n< L/2$.  
Then, if we use a transformation $V=(U-n)/(L-2n)$ from $U$ to $V$, 
$V=0$ corresponds to $U=n$ and $V=1$ to $U=L-n$.  
We can easily show that evolution equation on $V$ is obtained 
by replacing $U$ by $V$ and $L$ by 1 in (\ref{EBCA2}).  
Equation (\ref{EBCA2}) with $L=1$ is FI model itself 
and it is two-value CA.  
Therefore, if all initial values of $U$ is restricted to $n$ or $L-n$, 
then $U$ at any site is always $n$ or $L-n$ and evolutional state 
is the same as that of FI model by replacing 0 by $n$ and 1 by $L-n$.
\par
  Above fact implies a shape of the fundamental diagram.  
Consider an arbitrary state of FI model, that is, (\ref{EBCA2}) with
$L=1$.  
Let us assume density is $\rho$ and average flow is $Q$ for that state.
We can obtain a corresponding state of (\ref{EBCA2}) 
with $L>0$ by replacing 0 by $n$ and 1 by $L-n$.  
Then, using (\ref{EBCA2}), we can easily show that density of that 
state is $(1-\frac{2n}{L})\rho + \frac{n}{L}$ and 
flow is $(1-\frac{2n}{L})Q + \frac{2n}{L}$.  
Moreover, we can also show that a fundamental diagram of 
FI model is exactly given by two segments of which end points 
are $(0,0)$, $(\frac{1}{3}, \frac{2}{3})$ and $(\frac{1}{3},
\frac{2}{3})$, 
$(1,0)$ respectively.  
Therefore, a diagram of EBCA2 with $L>0$ has at least 
a segment with end points $(\frac{n}{L},\frac{2n}{L})$ 
and $(\frac{1}{3}+\frac{n}{3L},\frac{2}{3}+\frac{2n}{3L})$ 
and that with $(\frac{1}{3}+\frac{n}{3L}, \frac{2}{3}+\frac{2n}{3L})$ 
and $(1-\frac{n}{L}, \frac{2n}{L})$ for any $n$ ($0\le n<L/2$).  
Figure \ref{fig6} is a $L=7$ case and the shape of diagram is 
strictly a superposition of all these segments.
\section{Stability of flow}
In this section, we investigate stability of the branches in 
fundamental diagram obtained in the previous section.  
First, we study a uniform state $\cdots 11111 \cdots$, 
which gives the maximum flow in each model. 
Let us define a weak perturbation by a perturbation 
changing a local state $11$ to $20$, 
and a strong perturbation by a perturbation changing $1111$ to $2200$.  
Both perturbations clearly do not change the density.  
These perturbations mean that a car or two cars 
in the uniform flow suddenly decrease their speed and 
consequently fully-occupied sites `2' appear.
\par
  Figures \ref{fig7} (a)--(c) show an instability of uniform flow 
by a weak perturbation in multi-value SlS, 
EBCA2 and EBCA1 models respectively.  
We can observe a stop-and-go wave propagating backward 
from the perturbed site in Fig.\ref{fig7} (a) (multi-value SlS) and 
(c) (EBCA1), while the wave does not appear in (b) (EBCA2).  
In Fig.\ref{fig7} (c), flow $Q$ decreases by the weak perturbation 
and transits from A to F in Fig.\ref{fig3}.  
Moreover, we see that the final steady state contains 
locally two states corresponding to B and C in Fig.\ref{fig3}.  
Flow of state B is 5/6 and that of C is 1/2, both states 
are in the phase one, and they give the maximum and 
minimum flow in that phase.  
Numerical results from various initial states with 
a weak perturbation shows that a state of higher density 
tends to give this type of extreme local states on 
a branch after some time steps.  
Moreover, if we give a strong perturbation of uniform 
state $\cdots111\cdots$ corresponding to A, 
it goes down directly to G(1/2,1/2) in the phase zero.  
Fig.\ref{fig7} (d) shows an instability due to a strong perturbation in 
EBCA1 model.  In this case, stop-and-go wave does not occur.  
Moreover, we see that the final state consists of two local 
states $\cdots 200200 \cdots$ and $\cdots 222222 \cdots$ 
corresponding to D and E in Fig.\ref{fig3}, 
which also give the maximum and minimum flow in the phase zero respectively.
\par
  Since a perturbed uniform state becomes a congested 
one which consists of states giving the maximum and 
the minimum flow in a phase, we can predict a length 
of the final congested bunch.  
Assume $\alpha$ denotes a ratio of length of final congested 
bunch to length of total sites.  Density of B, F and C 
are 5/12, 1/2 and 3/4 respectively.  Thus we obtain
$$
  \frac{5}{12}(1-\alpha) + \frac{3}{4}\alpha = \frac{1}{2},
$$
and $\alpha$ becomes $1/4$ which coincides with that of Fig.\ref{fig7}
(c).  
In the case of the phase zero, we also obtain $\alpha=1/4$(Fig.\ref{fig7}(d)).
\par
  Contrary to above facts, there also exist some states on 
branches stable against a weak perturbation.  
Let us consider states on branch A--D in Fig.\ref{fig3}.  
There exists a state $\cdots 011011011 \cdots$ on the 
branch and we obtain $\cdots011020011\cdots$ by a weak perturbation.  
Against the perturbation, perturbed state is stable and it is still 
on the branch A--D.  We call this type of state metastable state.  
Moreover, let us consider a state $\cdots 111120111120 \cdots$ 
corresponding to F in Fig.\ref{fig3}.  
We can obtain $\cdots 201120111120 \cdots$ by a weak perturbation, 
and its flow does not change.  
Therefore, metastable states exist not only on branch A--D 
but also on the phase one branch.
In these cases, the effect of the site ``2'' does not spread over the
entire system during the time evolution.
Thus some part of the multiple branches are metastable states,
which can be observed in long time simulations 
starting from random initial conditions.
\section{Concluding discussions}
In this paper, some multi-value generalizations of rule--184 family 
are studied by using max--plus representation.  
We obtain evolution equations in a conserved form giving $(L+1)$-value
CA.  
We have proved that EBCA1 model is a generalization of SlS model, and it
gives multiple states and stop-and-go waves.  
We consider that a suitable 
traffic model gives those two phenomena at least.  
If we observe a real traffic data like Fig.\ref{fig1}, 
existence of multiple states is clear.  
In a real traffic, cars are considered to be always perturbed 
by some traffic effects.  
Therefore, we consider existence of stable branches 
against a weak perturbation is also desirable for a model.  
Moreover, stop-and-go waves are often observed in a real congested
traffic.  
Considering these points, EBCA1 is the most suitable model among 
the models of this paper.
\par
  Next, we give a presumption on a fundamental diagram of real data.
In Fig.\ref{fig1}, we see the multiple state clearly  
in an acceleration lane.
In a driving lane, since the average speed of cars is slower than the 
acceleration lane, flow tends to be stable and the branch is not clear.
In the acceleration lane, drivers move faster than cars 
in the driving lane, then 
over-dense free flow will be likely to occur and we can see the
multiple state around a critical density in the diagram.
Moreover, from the stability analysis we find that there are
metastable states even in the congested state in the EBCA1 model.
Although we cannot clearly see the fact due to 
fluctuation of data in Fig.1,
we expect that there exists both a metastable ``weak jam''(line B--C) 
and a ``strong jam''(line D--E) in a real highway traffic.
\par
  Finally, we point out future problems.  
We have investigated mainly about models with $L=2$. 
About multiple branches due to larger $L$, 
only EBCA2 gives clear branches and we can easily discuss their
stability.  
Though other models can also make multiple branches, 
transformation of perturbed states is complicated and 
its analysis is difficult.  We should solve this difficulty 
because we consider EBCA1 model is more suitable for a traffic model.  
Moreover, two-dimensional extension of these models and 
its application to a real traffic network is one of the 
important future problems.\\
\\\\
\centerline{\bf {Acknowledgment}}
The authors are grateful to Professors, Shinji Takesue, 
Hisao Hayakawa, Makoto
Kikuchi, Shinichi Tadaki, Yuki Sugiyama for fruitful 
discussions and helpful comments.  This work is partially 
supported by Grant-in-Aid from the Ministry of Education, Science and Culture.
\newpage

\newpage
\begin{flushleft}
{\large {\bf Appendix}} 
\end{flushleft}
In this appendix, we show that BCA model with $L=2$ can be 
interpreted as a two-lane model of rule--184 CA.  
Let us call one of the two lanes A-lane and the other B-lane.  
Both lanes are divided into discrete sites as shown in Fig.A1.  
Assume $A_j^t$ and $B_j^t$ denote number of cars 
at site $j$ and time $t$ in A-lane and B-lane respectively.  
Since all sites can hold only one car at most, $A_j^t$ and $B_j^t$ 
are always 0 or 1.  A car at site $j$ in A-lane 
moves according to the following rule:
\begin{itemize}
\item[(a)] If site $j+1$ in A-lane is empty, the car moves to that site.
\item[(b)] If site $j+1$ in A-lane is not empty and site $j$ 
and $j+1$ in B-lane are both empty, the car can move to site $j+1$ in B-lane.
\item[(c)] Otherwise, the car stays at site $j$ in A-lane.
\end{itemize}
As for a car in B-lane, symmetrical rule with respect to lane symbol is
applied.  
The above rule can be interpreted as follows: 
Every car move in its own lane prior to the other lane (rule (a)).  
However, if a car can not move in its own lane and only if 
there are no traffic in the other lane, 
it changes a lane and move forward in the other lane (rule (b)).  
Since cars move independently in each lane according to rule--184 
if rule (b) is omitted, the above rule can be interpreted as 
a two-lane model based on rule--184 with lane-changing effect.
\par
  We can express the rule by a couple of evolution equations as follows:
\begin{eqnarray}
 A^{t+1}_j &=& A^t_j + \min(A^t_{j-1},1-A^t_j)
                     - \min(A^t_j,1-A^t_{j+1}) \nonumber\\
           & & + \min(1-A^t_{j-1},1-A^t_j,B^t_{j-1},B^t_j)
               -\min(A^t_{j},A^t_{j+1},1-B^t_{j},1-B^t_{j+1}),  \label{A1} \\
 B^{t+1}_j &=& B^t_j + \min(B^t_{j-1},1-B^t_j)
                     - \min(B^t_j,1-B^t_{j+1}) \nonumber\\
           & & + \min(1-B^t_{j-1},1-B^t_j,A^t_{j-1},A^t_j)
               - \min(B^t_{j},B^t_{j+1},1-A^t_{j},1-A^t_{j+1}). \label{B1}
\end{eqnarray}
The last two terms in both equations express lane-changing effect.  
If those terms are omitted, both equations become independent 
each other and are equivalent to rule--184 CA (BCA model with $L=1$).  
If $U_j^t$ is defined by
\begin{equation}
 U^t_j=A^t_j+B^t_j,
\end{equation}
it denotes a sum of cars of both lanes.  
Since $A_j^t$ and $B_j^t$ is always 0 or 1, 
we can derive an evolution equation on $U_j^t$ 
from (\ref{A1}) and (\ref{B1}) as follows:
\begin{equation}
  U_j^{t+1} = U_j^t + \min(U_{j-1}^t,\,2-U_j^t) - \min(U_j^t,\,2-U_{j+1}^t).
\end{equation}
This equation is equivalent to BCA with $L=2$.
\newpage
\centerline{\bf {Figure Captions}} 
\def\labelenumi{Fig.\theenumi}
\begin{enumerate}
\item  \label{fig1}
Observed data of flow (vehicles$/$5min.) versus density (vehicles$/$Km) on Tomei
       expressway.  
This diagram was taken by Japan Highway Public Corporation. 
(a) driving lane, (b) acceleration lane.
\item  \label{fig2}
Fundamental diagrams of multi-value CA models with $L=2$ and $K=30$: 
(a) multi-value QS, (b) multi-value SlS, (c) EBCA2, (d)EBCA1 models. 
 \item \label{fig3}
Schematic fundamental diagram of EBCA1 model.
 \item \label{fig4}
Time evolution of flow in EBCA1 model. (a) $K=60$, (b) $K=240$, 
(c) power spectrum $I$ versus frequency $f$ of traffic flow for $K=240$.
 \item \label{fig5}
Fundamental diagram of EBCA1 model with $L=1$.
 \item \label{fig6}
Fundamental diagram of EBCA2 model with $L=7$.
 \item \label{fig7}
Instability of uniform flow due to a weak perturbation. (a) multi-value
       SlS, 
(b) EBCA2, (c) EBCA1 models. 
(d) shows an instability due to a strong perturbation in EBCA1 model.  
In all figures, black, dark gray, light gray squares denote a value 
2, 1, 0 respectively.
 \item[Fig.~A1]
Interpretation of BCA with $L=2$ as a model of coupled single lanes.
\end{enumerate}
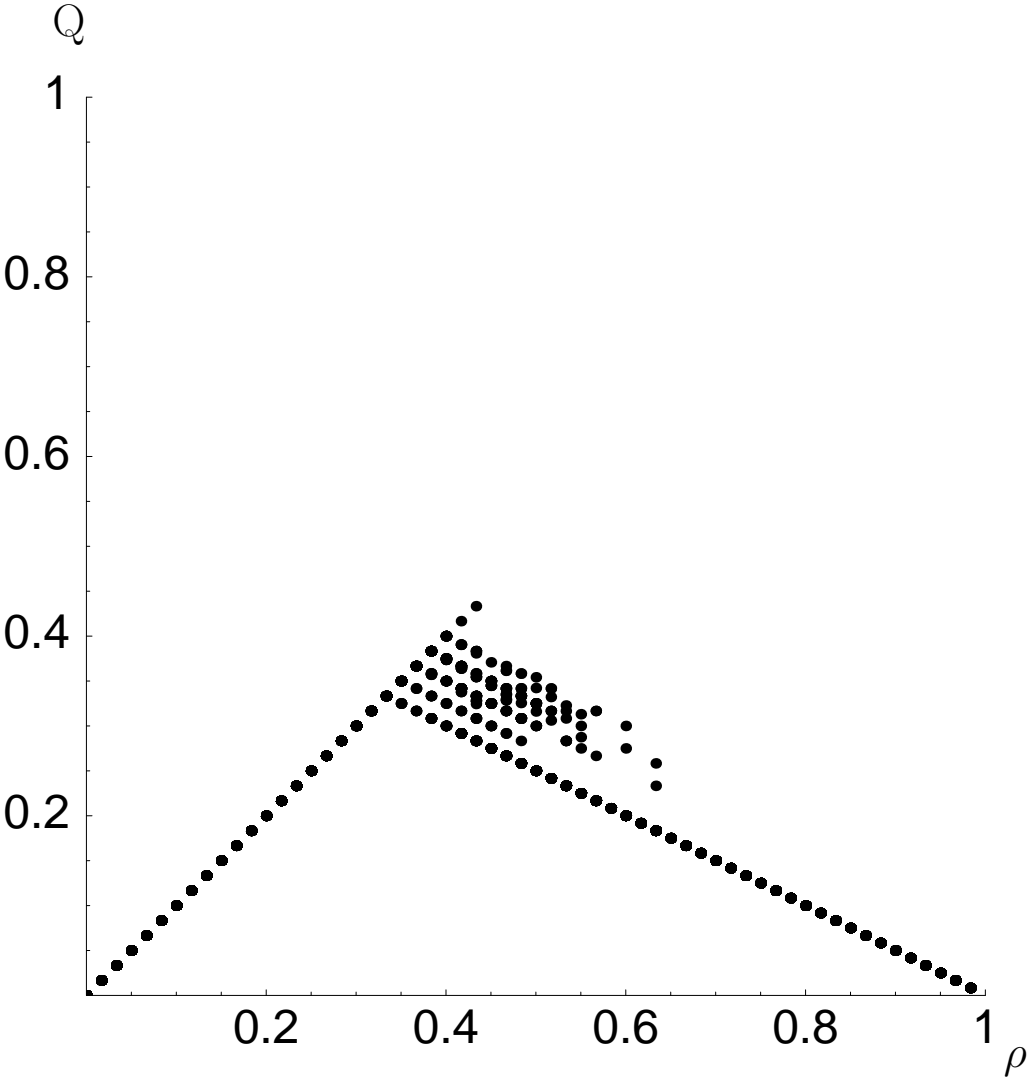
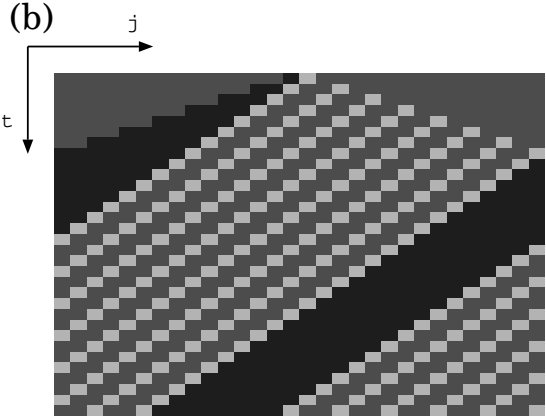
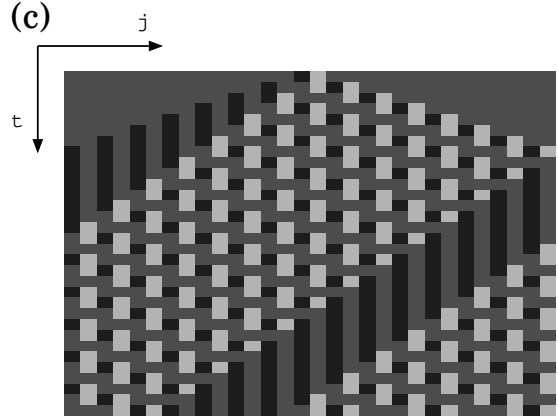
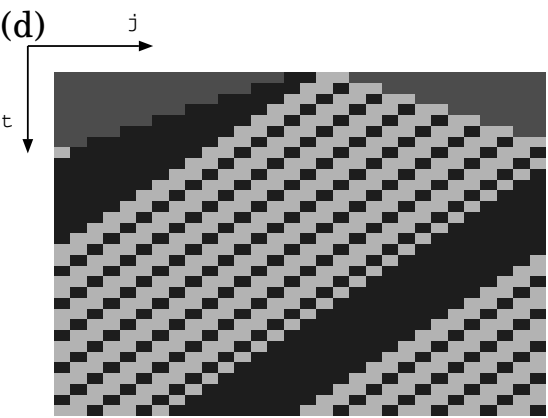
\end{document}